\documentclass[english,aps,twocolumn]{revtex4}
\usepackage[T1]{fontenc}
\usepackage[latin9]{inputenc}
\usepackage{textcomp}
\usepackage{amsmath}
\usepackage{amssymb}
\usepackage{graphicx}
\usepackage{esint}

\makeatletter
\@ifundefined{textcolor}{}
{%
 \definecolor{BLACK}{gray}{0}
 \definecolor{WHITE}{gray}{1}
 \definecolor{RED}{rgb}{1,0,0}
 \definecolor{GREEN}{rgb}{0,1,0}
 \definecolor{BLUE}{rgb}{0,0,1}
 \definecolor{CYAN}{cmyk}{1,0,0,0}
 \definecolor{MAGENTA}{cmyk}{0,1,0,0}
 \definecolor{YELLOW}{cmyk}{0,0,1,0}
}

\makeatother

\usepackage{babel}
\begin{document}

\title{High-gain weakly nonlinear flux-modulated Josephson parametric amplifier
using a SQUID-array}

\author{X. Zhou$^{1}$, V. Schmitt$^{1},$ P. Bertet$^{1},$ D. Vion$^{1}$,
W. Wustmann$^{2}$, V. Shumeiko$^{2}$, and D. Esteve$^{1}$}

\affiliation{$^{1}$Service de Physique de l'Etat Condens{é}/IRAMIS/DSM (CNRS
URA 2464), CEA Saclay, 91191 Gif-sur-Yvette, France}

\affiliation{$^{2}$Chalmers University of Technology, S-41296 Goteborg, Sweden. }

\pacs{85.25.\textminus{}j, 84.40.Dc, 42.50.Lc, 42.65.Yj}

\date{\today}
\begin{abstract}
We have developed and measured a high-gain quantum-limited microwave
parametric amplifier based on a superconducting lumped LC resonator
with the inductor L including an array of 8 superconducting quantum
interference devices (SQUIDs). This amplifier is parametrically pumped
by modulating the flux threading the SQUIDs at twice the resonator
frequency. Around 5 GHz, a maximum gain of 31 dB, a product amplitude-gain
$\times$ bandwidth above 60 MHz, and a 1 dB compression point of
-123 dBm at 20 dB gain are obtained in the non-degenerate mode of
operation. Phase sensitive amplification-deamplification is also measured
in the degenerate mode and yields a maximum gain of 37 dB. The compression
point obtained is 18 dB above what would be obtained with a single
SQUID of the same inductance, due to the smaller nonlinearity of the
SQUID array.
\end{abstract}
\maketitle
Although superconducting parametric amplifiers based on Josephson
junctions are known and understood for decades \cite{Barone,Yurke and Buks},
they have recently received an increased attention \cite{Lenhert}
because of their ability to measure single quantum objects and engineer
quantum fluctuations of a microwave field. They are extensively used
to readout superconducting quantum bits \cite{Devoret JPC,ParamFluxQ,quantum trajectories}
or mechanical resonators \cite{Teufel} at or near the quantum limit,
i.e with minimum back-action imposed by quantum mechanics for the
given amount of information taken on the system. They permitted for
instance the measurement of quantum trajectories \cite{quantum trajectories}
and the implementation of quantum feedback schemes \cite{Siddiqi Quantuum Feedback,Reversing quantum trajectories}.
In the field of quantum microwaves, they are also used to squeeze
quantum noise and produce itinerant squeezed states for encoding quantum
information \cite{Squeezed noise Mallet,Gross squeezing} or demonstrating
fundamental effects like the reduction of the radiative decay of an
artificial atom \cite{Siddiqi atom relaxation in squeezed}.

Compared to the noisier high electron mobility transistor (HEMT) based
amplifiers, these Josephson parametric amplifiers (JPA) suffer from
limited bandwidth and from gain saturation at extremely low input
power. A strong effort is thus made to increase the bandwidth and
to mitigate saturation of JPAs by varying their design and mode of
operation \cite{siddiqi PRB,highdynamc,JohnMartinis,wifi paramp}.
In all cases, parametric amplification of a signal at angular frequency
$\omega_{S}$ occurs by transfer of energy from a pump at frequency
$\omega_{P}$ to the signal and to a complementary idler frequency
$\omega_{I}$. For amplifiers based on resonators, one distinguishes
the case of intrinsically nonlinear resonators with bare frequency
$\omega_{R}$ that are pumped at $\omega_{P}\simeq\omega_{S}\simeq\omega_{R}$
directly on their signal line, from the (possibly linear) resonators
whose frequency $\omega_{R}$ is parametrically modulated with a pump
tone at $\omega_{P}\simeq2\omega_{S}\simeq2\omega_{R}$ on a dedicated
line separated from the signal port. In the first case the intrinsic
nonlinearity of the resonator is usually obtained by implementing
all or part of its inductance by Josephson junctions (or superconducting
weak links). The pumping at $\omega_{P}$ at sufficiently high amplitude
modulates this nonlinear inductance at $2\omega_{P}$, and is responsible
for a 4-wave mixing such that $2\omega_{P}=\omega_{S}+\omega_{I}$.
In the present work, we are interested in the second case \cite{Wilson parametric oscillator,Paramp NEC},
for which the nonlinearity is due to an externally imposed parametric
modulation of the frequency and is responsible for a 3-wave mixing
such that $\omega_{P}=\omega_{S}+\omega_{I}$. The interest of this
3 wave mixing is that no pump mode propagates along the input and
output signal lines and can blind a detector or spoil a squeezed field,
at a close frequency. In practice, the true parametric modulation
is usually obtained by embedding in the resonator inductance one or
several SQUIDs, the Josephson inductance of which is modulated by
an ac magnetic flux. The nonlinearity of the resonator inherited from
the SQUID(s) is in this case an unwanted feature, which leads to saturation
of amplification, and should thus be kept low. So besides the advantage
of getting rid of the pump along the signal lines, a truly parametrically
pumped amplifier can also be made more robust against saturation by
reducing its nonlinearity without having to pump it more strongly.
In this work, we test this idea and demonstrate a weakly nonlinear
JPA with high gain, made of a lumped LC resonator with the inductor
$L$ terminated by a SQUID array. The manuscript first summarizes
the theoretical description of such a JPA, then describes the device
implemented and its characterization setup, and finally presents the
experimental data and a comparison between measured and calculated
gain, bandwidth, and saturation.

\section{Theoretical summary and design choices\label{sec:Theoretical-summary}}

The specificity of the JPA presented here (pure 3 wave-mixing with
spurious nonlinearity) makes the standard classical description of
parametric amplifiers \cite{Yurke and Buks} not directly applicable
to it. This is why a comprehensive theoretical summary is given here,
based on the theoretical work \cite{Paramp Vitaly} (note that a similar
theoretical treatment can be found in \cite{Eichler-Wallraff}). The
equivalent circuit of the JPA is shown in the bottom-right corner
of Fig. \ref{fig:setup}b. For a DC flux bias $\Phi_{DC}$, a parametric
modulation $\delta L_{A}\cos(\omega_{P}t)$ of its array inductance
$L_{A}$, and a microwave input signal $V_{S}\cos(\omega_{S}t+\chi)$,
the JPA equation of motion at the lowest nonlinear order in phase
$\varphi=1/\varphi_{0}\int Vdt$ across the total inductance $L=L_{g}+L_{A}(\Phi_{DC})$
or the capacitance $C_{R}$ is

\begin{equation}
\varphi''+2\Gamma_{a}\varphi'+\omega_{R}^{2}[1+a_{P}(\Phi_{AC})\cos(\omega_{P}t)]\varphi+\alpha_{1}\varphi^{3}=\varphi_{S}\cos(\omega_{S}t+\chi),
\end{equation}

\noindent where $\varphi_{0}=\hbar/2e=\Phi_{0}/2\pi$ is the reduced
flux quantum, $\omega_{R}=2\pi f_{R}=1/\sqrt{L(C_{R}+C_{c})}$ the
frequency of the resonator, $\omega_{P}=2\pi f_{P}$ the pumping frequency,
$\Gamma_{a}=\omega_{R}/2Q$ its amplitude decay rate, $a_{P}=\delta L_{A}(\Phi_{AC})/L$
the relative pumping amplitude, $\alpha_{1}=-\omega_{R}^{2}p^{3}/2N^{2}$
the Josephson nonlinearity coefficient with $N$ the number of SQUIDs
and $p=L_{J}/L$ the so-called participation ratio of the total Josephson
inductance $L_{J}$ to the total inductance $L$; $\varphi_{S}$ is
the drive amplitude proportional to $V_{S}$. Taking into account
the finite ratio $\beta$ of each SQUID loop inductance to the inductance
$L_{J1}$ of a single junction, the SQUID array inductance is $L_{A}=NL_{J1}\beta/4+L_{J}$
with $L_{J}=NL_{J1}/\left\{ 2\left[\cos\left(x\right)+\beta/2\sin^{2}\left(x\right)\right]\right\} $
\cite{Agustin Squid inductance} and $x=\pi\Phi/\Phi_{0}$.

Equation (1) contains the parametric nonlinearity $\cos(\omega_{P}t)\varphi$
and the intrinsic Josephson nonlinearity $\alpha_{1}\varphi^{3}$
mentioned in the introduction. We rewrite it in the frame rotating
at $\omega_{P}/2$ using the slow complex internal amplitude $A(t)$
defined by $\varphi(t)=A(t)\sqrt{Z_{R}}\mathrm{\mathrm{e}^{-i\omega_{P}t/2}}+\mathrm{cc}$,
as well as the constant complex amplitude $B_{S}$ of the input signal
and the slow output amplitude $C(t)$ related to the input and output
voltages $V_{in,out}$ by $V_{in}(t)=i\varphi_{0}\sqrt{Z_{0}\omega_{S}}B_{S}\mathrm{\mathrm{e}^{-i\omega_{S}t}}/2+\mathrm{cc}$
and $V_{out}(t)=i\varphi_{0}\sqrt{Z_{0}\omega_{S}}C(t)\mathrm{\mathrm{e}^{-i\omega_{S}t}}/2+\mathrm{cc}$.
Here, $Z_{0}$ is the impedance of the line, $Z_{R}=\sqrt{L/(C_{R}+C_{c})}$
the characteristic impedance of the resonator, and $\mathrm{cc}$
stands for the complex conjugate of the previous term. Neglecting
fast oscillating terms (rotating wave approximation), one obtains
\cite{Paramp Vitaly}

\begin{equation}
\begin{cases}
iA'+(i\Gamma_{a}+\delta+\alpha\left|A^{2}\right|)A+\epsilon A^{*}=\sqrt{2\Gamma_{a}}B_{S}\mathrm{\mathrm{e}^{-i\Delta t}}\\
C=-i\sqrt{2\Gamma_{a}}A+B_{S}
\end{cases},
\end{equation}

\noindent where $\delta=\omega_{P}/2-\omega_{R}$ is the pump to resonator
detuning, $\Delta=\omega_{S}-\omega_{P}/2$ the signal to pump detuning,
$\alpha=-p^{3}Z_{R}\omega_{R}/16N^{2}$ the new nonlinear coefficient,
and $\epsilon=\omega_{R}a_{P}=2\omega_{R}\kappa\Phi_{AC}/\Phi_{0}$
the pumping strength with $\kappa\propto pQ$ the relative frequency
change per flux quantum deduced from the slope of the modulation curve
$\omega_{R}(\Phi_{DC})$.

Although the most general stationary solution of Eq. (2) is a sum
$\sum_{k\in\mathbb{Z}}\mathrm{A_{k}e}^{-ik\Delta t}$ of all harmonics
at frequencies $\omega_{P}/2+k\Delta$, only the signal $A_{S}=A_{1}$
and the idler $A_{I}=A_{-1}$ contributions happen to be non negligible
at not too high pumping strength. In this case, they obey

\begin{equation}
\begin{cases}
\left\{ \left[\delta_{r}+\alpha_{r}\left(\left|A_{S}\right|^{2}+2\left|A_{I}\right|^{2}\right)\right]+\Delta_{r}+i\right\} A_{S}+\epsilon_{r}A_{I}^{*}=\sqrt{2/\Gamma_{a}}B_{S}\\
\left\{ \left[\delta_{r}+\alpha_{r}\left(2\left|A_{S}\right|^{2}+\left|A_{I}\right|^{2}\right)\right]-\Delta_{r}+i\right\} A_{I}+\epsilon_{r}A_{S}^{*}=0\\
C_{S}=-i\sqrt{2\Gamma_{a}}A_{S}+B_{S}\\
C_{I}=-i\sqrt{2\Gamma_{a}}A_{I}
\end{cases}
\end{equation}

\noindent with $\delta_{r}=\delta/\Gamma_{a}$, $\Delta_{r}=\Delta/\Gamma_{a}$,
$\alpha_{r}=\alpha/\Gamma_{a}=-p^{3}Z_{R}Q/8N^{2}$ and $\epsilon_{r}=\epsilon/\Gamma_{a}$
the dimensionless detunings, nonlinear coefficient, and pumping strength,
respectively.

Our goal is to make the nonlinearity $\alpha_{r}$ as small as possible
and benefit from the linear signal and idler complex gains given by
system (3) when $\alpha_{r}=0$, i.e.

\begin{equation}
\begin{cases}
G_{S}=\frac{C_{S}}{B_{S}}=\frac{\delta_{r}^{2}-\Delta_{r}^{2}-1-\epsilon_{r}^{2}-2i\delta_{r}}{\delta_{r}^{2}-\Delta_{r}^{2}+1-\epsilon_{r}^{2}-2i\Delta_{r}}\\
G_{I}=\frac{C_{I}}{B_{S}}=\frac{2i\epsilon_{r}}{\delta_{r}^{2}-\Delta_{r}^{2}+1-\epsilon_{r}^{2}+2i\Delta_{r}}\mathrm{e}^{i2\chi}
\end{cases},
\end{equation}
yielding the power gains 

\begin{equation}
\left|G_{S}\right|^{2}=1+\left|G_{I}\right|^{2}=1+\frac{4\epsilon_{r}^{2}}{\left[1-\epsilon_{r}^{2}+\delta_{r}^{2}-\Delta_{r}^{2}\right]^{2}+4\Delta_{r}^{2}}.
\end{equation}

In the degenerate case corresponding to $\Delta=0$, the signal power
gain becomes phase $\chi$ dependent and is given by

\begin{equation}
\left|G_{S,deg}\right|^{2}(\chi)=1+4\epsilon_{r}\frac{2\left[\epsilon_{r}-\delta_{r}\cos(2\chi)\right]+(1+\epsilon_{r}^{2}-\delta_{r}^{2})\sin(2\chi)}{\left(1-\epsilon_{r}^{2}+\delta_{r}^{2}\right){}^{2}}.
\end{equation}
Equations (4-6) are valid only below the onset of parametric oscillations,
that is of pump-induced auto-oscillations at zero signal $B_{S}$
for $\epsilon_{r}>1+$$\delta_{r}^{2}$. For sufficiently large pumping
strength $\epsilon_{r}>0.42$ the power gain $\left|G_{S}\right|^{2}$
is larger than 2 at small $\Delta_{r}$ and $\delta_{r}$, and a gain
bandwidth $\Delta\omega=2\pi\Delta f$ at $-3\,\mathrm{dB}$ can be
defined. For the optimal pumping frequency $\delta_{r}=0$ we find

\begin{equation}
\frac{\Delta\omega}{2\Gamma_{a}}=\sqrt{\left(1+\epsilon_{r}^{2}\right)\left(\frac{2\epsilon_{r}}{\sqrt{\epsilon_{r}^{4}-6\epsilon_{r}^{2}-1}}-1\right)},
\end{equation}

\noindent which yields a gain bandwidth product
\begin{equation}
\left|G_{S}\right|\Delta\omega/2\Gamma_{a}\simeq1
\end{equation}
that is constant within 10\% above $7\,\mathrm{dB}$ gain. Then, saturation
can be evaluated approximately in an easy way by noticing that as
the internal amplitudes of oscillation $A_{S}$ and $A_{I}$ increase
with the pumping strength and gain, they tend to the same value when
$\left|G_{S}\right|^{2}\simeq\left|G_{I}\right|^{2}\gg1$ (see Eq.
(5)). Consequently, the terms in $\alpha_{r}$ in the first two equations
of system (3) also converge to close values and play the very same
role as the pump to resonator detuning $\delta_{r}$, which is itself
responsible for a gain drop given by Eq. (6). Equating $\alpha_{r}\left|A_{S}^{2}+2A_{I}^{2}\right|$
at $\delta_{r}=0$ to the value $\delta_{r,sat}=0.35\sqrt{1-\epsilon_{r}^{2}}$
that produces a $-1\,\mathrm{dB}$ drop of $\left|G_{S}\right|^{2}-1$,
leads to the following equivalent values for $A_{S}$ and $B_{S}$
(so called $1\,\mathrm{dB}$ compression point):
\begin{equation}
\begin{cases}
A_{S,sat}^{2}\simeq\frac{0.35}{\left|\alpha_{R}\right|}\frac{\sqrt{1-\epsilon_{r}^{2}}}{1+2\epsilon_{r}^{2}}\\
\frac{B_{S,sat}^{2}}{\Gamma_{a}}\simeq\frac{0.17}{\left|\alpha_{R}\right|}\frac{\left(1-\epsilon_{r}^{2}\right)^{5/2}}{1+2\epsilon_{r}^{2}}
\end{cases}
\end{equation}

In addition, saturation at large gain $\left|G_{S}\right|$ has to
occur when the peak current $i$ in the junctions is still well below
their critical current $i_{c}$. Since at $\delta_{r}\sim0$, $i/i_{c}=(1+\epsilon_{r})\left|A_{S}\right|p\sqrt{Z_{r}}/N$,
keeping $i/i_{c}<0.5$ yields the design rule

\noindent 
\begin{equation}
pQ>21/\sqrt{\left|G_{S,max}\right|+1},
\end{equation}

\noindent which imposes a minimum $p$ for low $Q$ and wide bandwidth
JPAs. 

In this work, we choose to implement a tunable amplifier in the 5-6
GHz range with a quality factor $Q$ of order 100, which should have
a product gain$\times$bandwidth of $\sim50\,\mathrm{MHz}$ according
to Eq. (8). To reduce the maximum microwave pumping power corresponding
to $\epsilon_{r}=Qa_{P}/2=1$, i.e. to a modulation $a_{P}\sim1\%$
of the total inductance, a high participation ratio $p\simeq0.5$
is chosen. On the other hand, in order to keep the nonlinearity $\alpha_{r}$
weak and to increase the $1\,\mathrm{dB}$ compression point $B_{S,sat}$,
the tunable inductance is implemented with $N=8$ SQUIDs. In this
case, the saturation power is increased by $N^{2}$ or $18\,\mathrm{dB}$,
compared to the case of a single SQUID with the same total Josephson
inductance. Finally, as $Z_{r}$ plays only a minor role in the nonlinearity$\alpha_{r}$
(in comparison with $p^{3}$and $N^{-2}$), its value will be simply
chosen at the best convenience for implementing the lumped element
resonator.

\section{Sample and measurement setup}

\begin{figure}[!t]
\includegraphics[width=8.5cm]{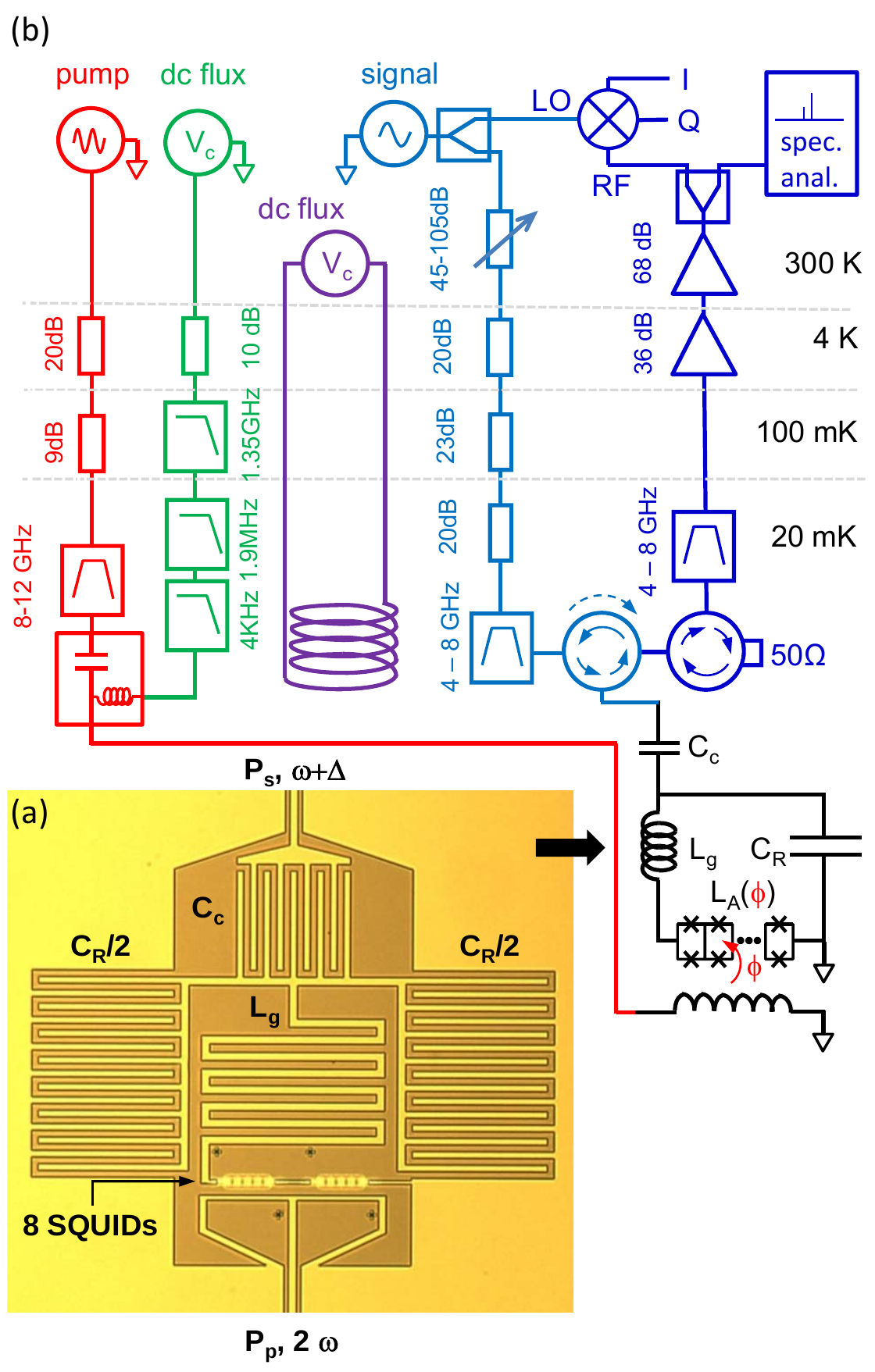} \caption{\label{fig:setup} Experimental setup. (a) Optical micrograph of the
tested parametric amplifier showing its $50\,\Omega$ coplanar waveguide
(CPW) signal input port (top), its coupling capacitor $C_{c}$, its
capacitor $C_{R}$ (left and right), its inductor $L_{g}$ (middle)
terminated by an 8-SQUID arrays with total inductance $L_{A}\left(\Phi\right)$,
and its split magnetic flux line coupled to a $50\,\Omega$ CPW (bottom).
The DC current in the flux line sets the dc flux $\Phi_{DC}$ and
the resonance frequency $f_{R}$ of the resonator, whereas the ac
current parametrically pumps the resonator at $\omega_{P}\simeq2\omega_{R}$.
The black arrow points to the equivalent circuit. (b) Electrical circuit
diagram showing (from left to right) the pump line, the dc flux line
added to the pump with a bias tee, an additional flux line feeding
a coil for compensating any flux offsets, the signal line, a circulator
routing the reflected and amplified signal to a measurement line through
an isolator protecting the sample from the noise of the first amplifier
placed at 4 K. Feeding lines are attenuated and filtered. The output
signal is split after amplification and analyzed both with a spectrum
analyzer and by homodyne demodulation.}
\end{figure}

An optical micrograph of the JPA and its equivalent circuit are shown
in Fig. \ref{fig:setup}a. This JPA is made of an interdigitated coplanar
capacitor to ground (split in two parts) with capacitance $C_{R}=0.40\,\mathrm{pF}$
, in parallel with an inductance to ground $L$ combining in series
a meander of inductance $L_{g}=0.80\,\mathrm{nH}$ with an array of
8 SQUIDs of total Josephson inductance$L_{J}=0.88\,\mathrm{nH}$ at
zero magnetic flux $\Phi$. Being designed to be operated in reflection,
this LC circuit is coupled to a single input-output signal line ($50\,\Omega$
coplanar waveguide - CPW ) through a $C_{c}=55\,\mathrm{fF}$ capacitance
yielding a characteristic impedance $Z_{r}\backsimeq65\,\Omega$ and
a quality factor $Q\simeq70$ at $\Phi=0$. On the other side of the
device a $50\,\Omega$ CPW line shorted to ground by two loops coupled
inductively to 4 SQUIDS each, serves both for their DC flux biasing
and for parametric pumping. Note that after compensation of any global
DC flux offset, the magnetic fluxes $\Phi$ are exactly opposite in
the left and right 4 SQUID sub-arrays, which yields the same inductance
modulation.

The device was fabricated on a thermally oxidized Si chip by sputtering
$170\,\mathrm{nm}$ of niobium and patterning the whole structure
(except the SQUID array) by optical lithography and $\mathrm{CF_{4}-Ar}$
reactive ion etching. The SQUID array was then fabricated by e-beam
lithography and double-angle evaporation of aluminum with oxidation
of the first Al layer. Each SQUID has a loop area of $8\,\mathrm{\mu m}\times15\,\mathrm{\mu m}$
and two junctions with nominal area $2.2\,\mathrm{\mu m}\times0.7\,\mathrm{\mu m}$
and tunnel resistance $141\,\Omega$, yielding $\beta\simeq0.1$.
The active antenna wires of the pump line are positioned $16\mathrm{\,\mu m}$
away from the SQUID centers.

The measurement set-up is schematized in Fig. \ref{fig:setup}b. A
small superconducting coil is used to compensate the global DC flux
offset. DC flux biasing and AC pumping of the SQUIDs are obtained
by two attenuated and filtered lines combined with a bias-tee. The
input line includes attenuators at various temperatures and a $4-8\,\mathrm{GHz}$
bandpass filter. The $-71.5\,\mathrm{dB}$ and $-51\,\mathrm{dB}$
transmissions of the input and pumping lines are calibrated with a
$\pm1\,\mathrm{dB}$ uncertainty. The reflected and amplified signal
is routed to the output line by a cryogenic circulator with $-18\,\mathrm{dB}$
isolation. This output line includes an isolator for protecting the
sample from higher temperature noise, a $4-8\,\mathrm{GHz}$ filter,
a cryogenic high electron mobility transistor (HEMT) amplifier at
4K with $38\,\mathrm{dB}$ gain and a calibrated noise temperature
of $3.8\,\mathrm{K}$, as well as additional room temperature amplifiers.
The output signal is finally analyzed using a spectrum analyzer or
a homodyne demodulator followed by a digitizer. Microwave generators
for the input signal and pump are precisely phase locked.

\section{Experimental results}

\begin{figure}
\includegraphics[width=8cm]{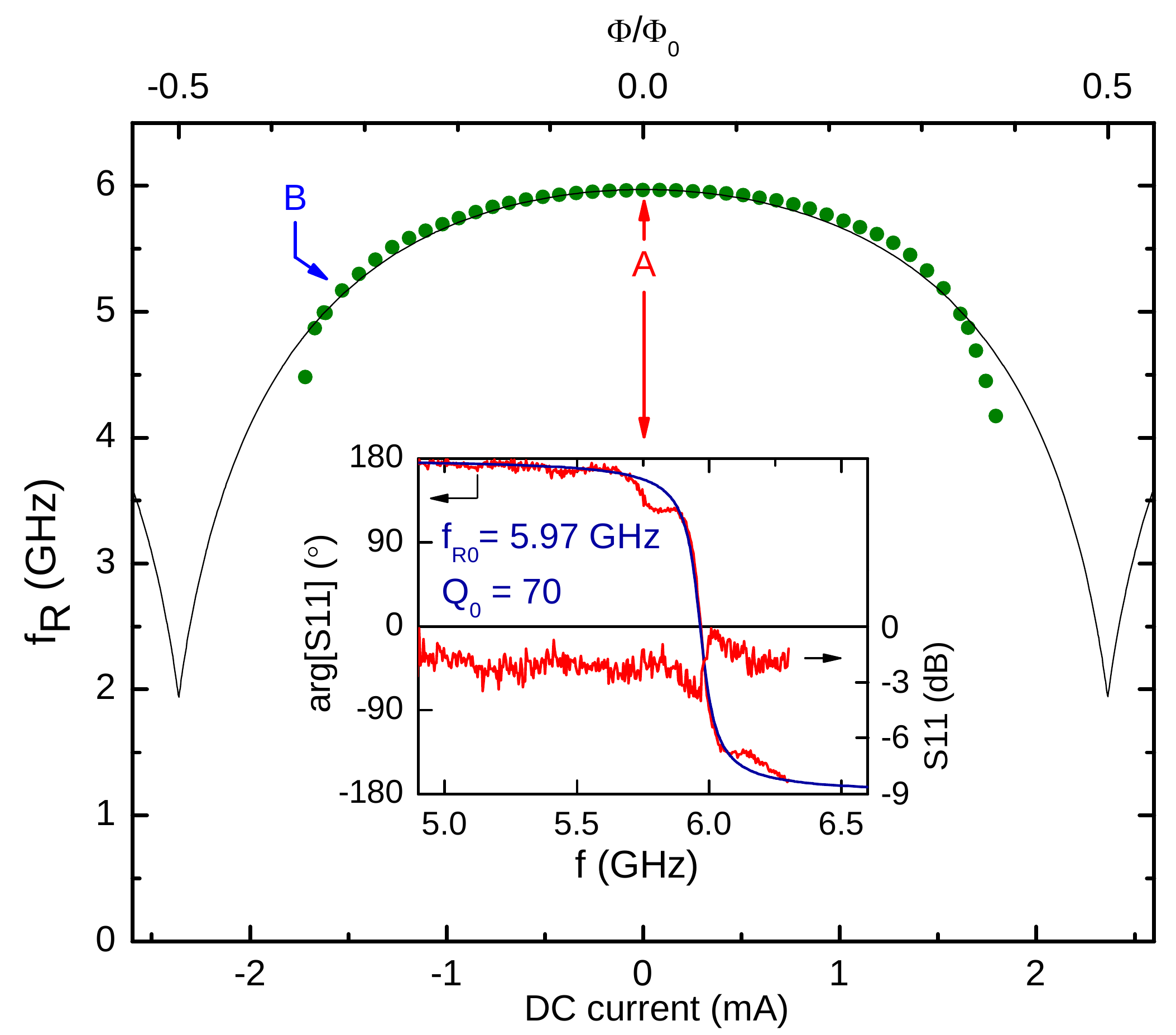} \caption{\label{fig:flux modulation} DC flux modulation. Experimental (dots)
and calculated (line) resonator frequency $f_{R}$ as a function of
the DC current in the on-chip flux line (see Fig.\ref{fig:setup}a)
measured by fitting the phase of a weak signal reflected on the resonator
in absence of parametric pumping, as shown in the inset for zero flux
bias (point A). We attribute the two shoulders on the sides of the
measured resonance to multiple wave interferences due to an imperfect
impedance matching somewhere in the setup. Parameters used for calculation
of the modulation curve are $f_{R0}$, $p=0.45$, and $\beta=0.1$
. Full characterization in the next figures are done at working point
B.}
\end{figure}

Measurements were performed in a dilution refrigerator at the temperature
of $30-40\,\mathrm{mK}$. As a preliminary characterization, the resonance
was measured with a vectorial network analyzer by recording the phase
of the reflected signal at zero pumping and at a nominal input power
$P_{S,n}=-126\,\mathrm{dBm}$ small enough to avoid any nonlinear
effects (all nominal powers mentioned here and below refer to powers
at the sample ports given the calibration of the lines). Inset of
Fig. \ref{fig:flux modulation} shows this resonance at zero flux
with a fit of the curve yielding a maximum frequency of $f_{R0}=5.97\,\mathrm{GHz}$.
The comparison with the $f_{R1}=8.06\,\mathrm{GHz}$ resonance frequency
of a similar resonator with shorted junctions yields $p=1-(f_{R0}/f_{R1})^{2}=0.45$,
close to the $0.42$ design value. Fitting the expression $-2\arctan[2Q(f_{S}/f_{R}-1)]$
to the measured resonance curve also gives the quality factor $Q_{0}\simeq70$,
with however limited accuracy due to a setup imperfection yielding
spurious multiple wave interferences (see shoulders in inset of Fig.
\ref{fig:flux modulation}). The main graph of Fig. \ref{fig:flux modulation}
shows the variation of $f_{R}$ as a function of the applied flux
$\Phi$ and its comparison with the theoretical prediction from section
I. The agreement is only qualitative especially above $0.35\,\Phi_{0}$
where $f_{R}$ decreases faster than predicted by our simple model
that does not include either the flux inhomogeneity in the different
SQUIDs or the possible penetration of the flux through the junctions.

For characterizing amplification, the input and output lines are then
connected as shown in Fig. \ref{fig:setup}. The signal and idler
gains are measured with the spectrum analyzer by comparing the output
powers of the signal and idler without and with parametric pumping
at $f_{P}=2f_{R}$. The gains increase with $\left|\Phi_{DC}\right|$
and the slope of the modulation curve at fixed absolute pumping power.
In the following measurement, signal amplification is fully characterized
at the working point ($\Phi_{1}/\Phi_{0}=-0.32$, $f_{R1}=5.17\,\mathrm{GHz}$),
i.e. point B on Fig. \ref{fig:flux modulation}, where the slope $\kappa_{1}=1.62$
is at the same time large and in agreement with the predicted value.
At this point the SQUID array model predicts a participation ratio
$p_{1}=0.59$ and a quality factor $Q_{1}=81$.

\begin{figure}
\includegraphics[clip,width=8.5cm]{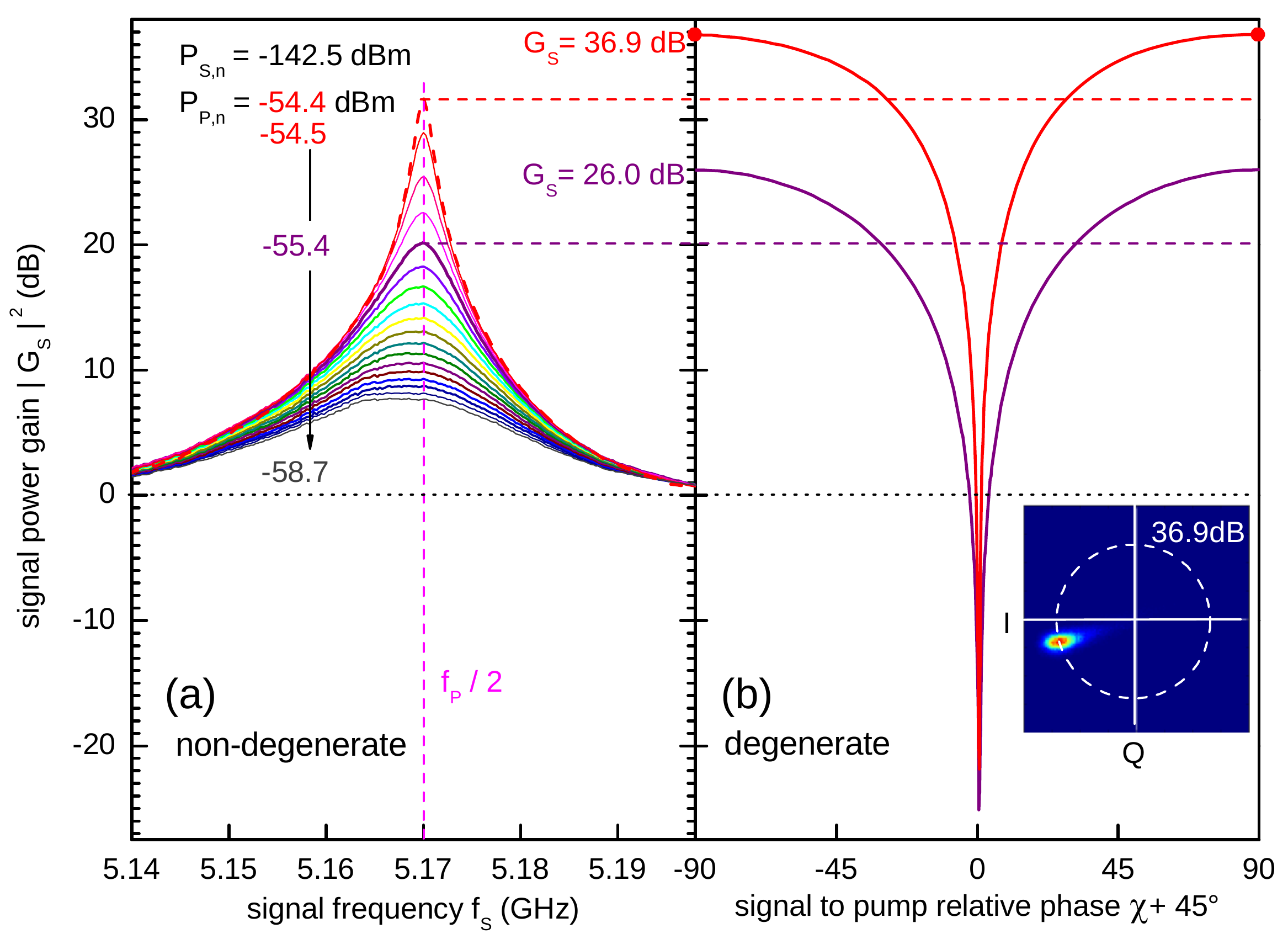} \caption{\label{fig:gain} Signal power gain $\left|G_{S}\right|^{2}$ at working
point B of Fig. \ref{fig:flux modulation} for a nominal input power
$P_{S,n}=-142.5\,\mathrm{dBm}$. (left) Non-degenerate gain as a function
of the signal frequency $f_{S}$ at different nominal pumping powers
$P_{P,n}$ between $-58.7\,\mathrm{dBm}$ and $-54.4\,\mathrm{\mathrm{dB}m}$
(top dashed curve just before the onset of parametric oscillation
in the absence of incident signal). (b) Phase sensitive degenerate
gain for $P_{P,n}=-54.4\,\mathrm{dBm}$ and $-55.4\mathrm{\, dBm}$.
(inset) Demodulated signal in the IQ plane at maximum gain ($P_{P,n}=-54.4\,\mathrm{dBm}$)
filtered at 1 MHz. I and Q voltages are digitized at 1 MSample/s during
2 s, and the color encodes the density of samples from 0 (dark blue)
to maximum (red).}
\end{figure}

\begin{figure}
\includegraphics[clip,width=8.5cm]{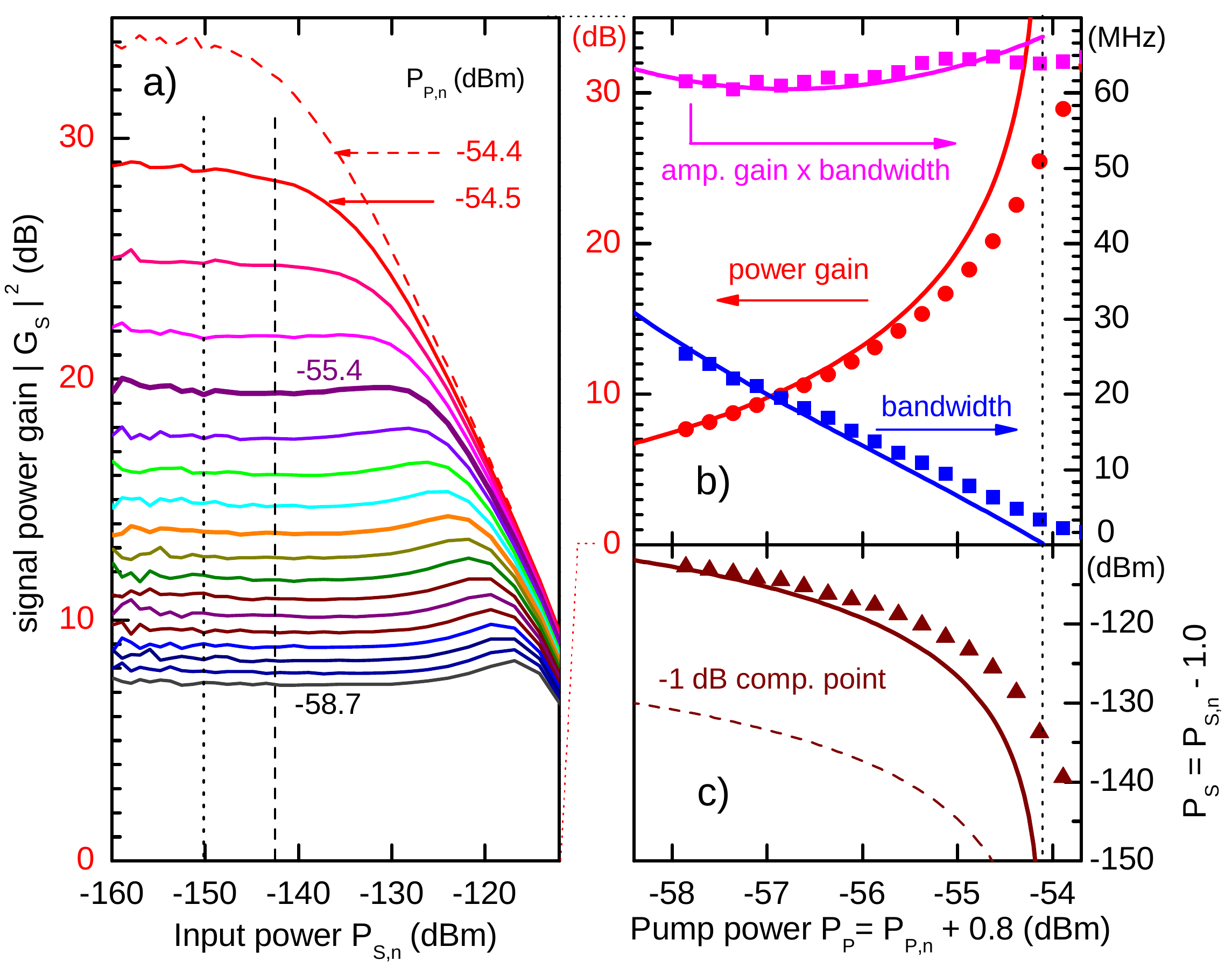} \caption{\label{fig:summary} Amplifier characterization at the working point
B of Fig. \ref{fig:flux modulation} for non-degenerate pumping. (a)
Signal power gain as a function of the nominal input power $P_{S,n}$
showing the saturation at the same nominal pumping powers $P_{P,n}$
as in Fig. \ref{fig:gain}a (top dashed line corresponds again to
the onset of parametric oscillation). Vertical dashed and dotted line
correspond to the input powers where gain was measured in Fig. \ref{fig:gain}
and where reference gain for saturation was defined, respectively.
(b) Power gain $\left|G_{S}\right|^{2}$, bandwidth $BW$, and product
$\left|G_{S}\right|\times BW$ deduced from measurements (dots) of
Fig. \ref{fig:gain}a at $P_{S,n}=-142.0\,\mathrm{dBm}$, and calculated
(solid lines) from the model with the parameters indicated in the
text. (c) 1~dB compression point deduced from (a) (dots), calculated
from the model (solid line), and calculated with the sames parameters
but only one SQUID (dashed line). Note that given the $\pm1\,\mathrm{dB}$
precision on the calibration of the pumping and signal lines, $+0.8\,\mathrm{dB}$
and $-1.0\,\mathrm{dB}$ were added to the nominal $P_{P,n}$ and
$P_{S,n}$ values to match the data to the theoretical curves at low
pumping strength. The vertical dotted line indicates the frontier
between parametric amplification and parametric oscillation (infinite
gain) for the linear model.}
\end{figure}

The non-degenerate ($\Delta\neq0$) signal power gain $\left|G_{S}\right|^{2}$
is measured with the pump frequency $f_{P}=2f_{R}$ ($\delta=0$)
as a function of the signal frequency $f_{S}$ for increasing nominal
pump power $P_{P,n}$, at an input power $P_{S,n}=-142.5\,\mathrm{dBm}$
sufficiently low to avoid the saturation at the highest gain. Close
to the resonance, a minimum detuning $\delta/2\pi=5\,\mathrm{kHz}$
is used to avoid operation in the degenerate mode. Figure \ref{fig:gain}a
shows the gain increase up to $31.8\,\mathrm{dB}$ (dashed top curve,
for which parametric oscillations are about to start) and the corresponding
bandwidth decrease. The maximum power gain $\left|G_{S}\right|^{2}$
and the corresponding $-3\,\mathrm{dB}$ bandwidth $\Delta f$ deduced
from Fig. \ref{fig:gain} are plotted on Fig. \ref{fig:summary}b
together with the amplitude-gain $\times$ bandwidth product $\left|G_{S}\right|\Delta f$.
This product happens to be almost constant around $61\,\mathrm{MHz}$
over the whole $7\,\mathrm{dB}-30\,\mathrm{dB}$ gain range. Besides,
the idler gain (data not shown) approaches the signal gain at large
values.

In order to check that the amplifier operates close to the quantum
limit, i.e. with a noise temperature of order $T_{N}=hf_{R}/2k_{B}\simeq125\,\mathrm{mK}$
\cite{noise review Clerk devoret}, the variation of the signal and
noise powers are compared when switching on and off the pump: from
the $2.9\,\mathrm{dB}$ increase of the noise when switching on a
$18.4\,\mathrm{dB}$ gain, from the calibrated $3.8\pm0.3\,\mathrm{K}$
noise temperature of the HEMT amplifier alone in a separate run, and
from the $1.7\pm0.2\,\mathrm{dB}$ attenuation of elements placed
below 250 mK between the sample and the HEMT amplifier, we deduce
an apparent noise temperature of only $80\pm10\,\mathrm{m\mathrm{K}}$.
This value is smaller than the expected quantum limit of $125\,\mathrm{mK}$,
a discrepancy that shows that modeling the line by a simple attenuator
is not sufficient, as supported by our observation of the setup imperfection
already mentioned. This result nevertheless indicates that our JPA
is not far from the quantum limit. A more precise determination of
$T_{N}$ would require a much more precise control and calibration
of the low temperature part of the measurement line, as well as a
switch to connect the detection chain either to the JPA or to a low
temperature reference noise source \cite{Squeezed noise Mallet}.

The phase dependent gain in the degenerate case ($\delta=0$ ) was
then measured with $f_{S}=5.17\,\mathrm{GHz}$ by varying the phase
$\chi$ of the signal with respect to the pump; it is shown on Fig.
\ref{fig:gain}b for the two values of the pump power that correspond
to a $20\,\mathrm{dB}$ gain and to the maximum gain in the non-degenerate
mode. As expected the maximum degenerate gain is $6\,\mathrm{dB}$
larger than the non-degenerate gain at almost the same frequency.
At the highest degenerate gain of $36.9\,\mathrm{dB}$, it was checked
using the IQ demodulator (see inset of Fig. \ref{fig:gain}b) that
the phase of the amplified signal is stable over minutes and that
the output signal drops down to zero (no parametric oscillation) when
the input signal is switched off. As the phase $\chi$ is varied,
the measured degenerate gain varies as expected, the lowest value
of $-25\,\mathrm{dB}$ resulting from the uncontrolled interference
between the deamplified signal and the $\sim-18\,\mathrm{dB}$ leak
of input signal through the circulator (see Fig. \ref{fig:setup}).
This strong deamplification and the low noise temperature indicate
that our JPA could also be used as a vacuum squeezer. In the inset
of Fig. \ref{fig:gain} the elongation of the Gaussian spot along
the amplified quadrature shows that after parametric amplification
the noise coming from the sample at $f_{S}=5.17\,\mathrm{GHz}$ overcomes
the noise of the cryogenic amplifier placed at $4\,\mathrm{K}$, the
size of which is given by the spot size in the perpendicular direction.
In this latter direction we observe that the spot size is reduced
by 1.1\% when switching on the parametric pumping. This reduction
is twice as small as the 2.2\% expected from deamplification of vacuum
noise, which is again related to the difficulty to determine the noise
temperature of the whole setup. 

Finally, the saturation of the JPA is measured by recording the non-degenerate
signal power gain as a function of the signal input power $P_{S,n}$
for the same series of pump powers $P_{P,n}$ as before (see Fig.
\ref{fig:summary}a). The signal gain is almost constant at low input
power and then decreases above a $P_{P,n}$ dependent threshold in
$P_{S,n}$ (however with a small bump of up to $1\,\mathrm{dB}$ just
before saturation possibly due to higher orders in non-linearity).
In practice, the $1\,\mathrm{dB}$ compression point is defined as
the input power $P_{S,sat}$ at which the gain is $1\,\mathrm{dB}$
lower than at $P_{S,n}=-150\mathrm{\, dBm}$; it is plotted on Fig.
\ref{fig:summary}c. The set of measurements of Figs. \ref{fig:summary}b-c
is then compared to the linear model of section \ref{sec:Theoretical-summary}:
the power gain, bandwidth, product amplitude-gain $\times$ bandwidth,
as well as the $1\,\mathrm{dB}$ compression point of Eqs. (6-9) are
calculated by using the values of $f_{R1}$, $p_{1}$, $\kappa_{1}$
and $Q_{1}$ indicated above and are plotted on Fig. \ref{fig:summary}.
Given the $\pm1\,\mathrm{dB}$ uncertainty on the calibration of the
signal and pump lines, the nominal input and pump powers were shifted
by $+0.8\,\mathrm{dB}$ and $-1.0\,\mathrm{dB}$ to match the theory
at the lowest pumping power. The agreement between the overall measured
data and the model is surprisingly good given the crudeness of the
linear model. This fair agreement validates the idea of increasing
the number $N$ of SQUIDs to increase the saturation power that scales
with $N^{2}$. With a single SQUID having the same total inductance
as the array implemented here (about $1.7\,\mathrm{nH}$), the saturation
would have been $N^{2}=18\,\mathrm{dB}$ lower, as indicated by the
dashed line of Fig. \ref{fig:summary}c. The discrepancy between experimental
data and the model increases with $P_{p}$ as the nonlinearity plays
a more important role, and the actual parametric amplification region
extends a bit over the theoretical parametric oscillation region of
the linear model (dotted line of Fig. \ref{fig:summary}).

The performances of the present device are comparable to those of
other truly parametric amplifiers recently made. Due to our choice
of a rather large $Q\thicksim70$, the gain bandwidth product is smaller
than what was obtained for instance in \cite{Hartridge-Siddiqi} with
$Q\thicksim10$. In \cite{JohnMartinis}, the direct coupling of the
resonator to a cleverly engineered frequency dependent external impedance
yielded an even lower Q and a bandwidth above 500 MHz. Despite the
use of $N=8$ SQUIDs, the 1dB compression point obtained here is not
very high due to its scaling as $N^{2}Q^{-2}p^{-3}$ and to the large
participation ratio and quality factors chosen to minimize the pump
power: it is however about 12 dB above a similar amplifier made of
a single SQUID with about the same critical current \cite{Paramp NEC}
and only a few dB below another one \cite{Hartridge-Siddiqi} with
smaller participation ratio $p$ (three times larger critical current
Ic) and Q.

In terms of perspectives, equations (8-10) predict that with a similar
geometry N$\sim10$, a smaller $Q\sim10$, and higher critical currents
yielding $p\sim0.25$, a bandwidth of $\sim50\,\mathrm{MHz}$, and
a compression point $\sim-100\,\mathrm{dBm}$ should be obtained at
$20\,\mathrm{dB}$ gain. This would require a larger pump power, i.e.
a larger flux modulation $\Phi_{AC}\varpropto1/pQ$ at constant gain,
which would reach $0.1\Phi_{0}$. Such a large modulation could be
technically difficult to achieve. Increasing the number of SQUIDs
is also an obvious optimization axis: If theoretically the array length
has just to be kept much smaller than the pump wavelength so that
all SQUIDS are pumped in phase, the practical difficulty is to DC
flux bias and homogeneously modulate all the SQUIDs.

In summary, a lumped element truly parametric Josephson amplifier
has been designed and characterized. Its inductance is implemented
by a SQUID array to limit its nonlinearity and increase the maximum
allowed input power. With a quality factor of 70~-~80, this simple
device provides a gain of up to $30\,\mathrm{dB}$, a product amplitude-gain$\times$bandwidth
of $61\,\mathrm{MHz}$, and a $1\,\mathrm{dB}$ compression point
of $-123\,\mathrm{dBm}$ at $20\,\mathrm{dB}$ gain. Although its
behavior is in agreement with theory and demonstrates the advantage
of using a SQUID array, it can still be optimized by reducing both
its quality factor and its Josephson participation ratio to the inductance
and/or by increasing the number of SQUIDs in the array. Operated close
to the quantum limit, this truly parametric amplifier could also be
used as a quiet and strong squeezer in degenerate mode or as the first
stage of amplification in a superconducting quantum bit readout.

\section*{Acknowledgment}

We gratefully acknowledge discussions within the Quantronics group,
technical support from P. Orfila, P. Senat, J.C. Tack and Dominique
Duet, as well as financial support from the European research contract
SCALEQIT.

\end{document}